\documentclass[doublecol]{epl2} 

\usepackage{amsmath}

\title{Bayesian inference of wall torques for active Brownian particles}
\shorttitle{Bayesian inference of wall torques for active Brownian particles} 

\author{S. Lambert \and M. Duchêne \and S. Klumpp}
\shortauthor{S. Lambert \etal}

\institute{                    
  University of Göttingen, Institute for the Dynamics of Complex Systems, Friedrich-Hund-Platz 1, 37077 Göttingen, Germany.
}

\abstract{
The motility of living things and synthetic self-propelled objects is often described using Active Brownian particles. To capture the interaction of these particles with their often complex environment, this model can be augmented with empirical forces or torques, for example, to describe their alignment with an obstacle or wall after a collision.
Here, we assess the quality of these empirical models by comparing their output predictions with trajectories of rod-shaped active particles that scatter sterically at a flat wall. We employ a classical least-squares method to evaluate the instantaneous torque. In addition, we lay out a Bayesian inference procedure to construct the posterior distribution of plausible model parameters. In contrast to the least squares fit, the Bayesian approach does not require orientational data of the active particle and can readily be applied to experimental tracking data.
}

\begin{document}

\maketitle

    

\section{Introduction}


Self-propelled particles are a widely studied type of active matter that includes a broad range of systems, from microbial motility to animal flocking \cite{romanczuk_active_2012, bechinger_active_2016, Zafeiris2012}. Active Brownian Particles (ABP) form one of the simplest models of such systems, based on a fixed active propulsion force and diffusion to describe the motion. These minimal descriptions are often extended to include interactions between the particles \cite{telezki_simulations_2020, khan_toward_2024}, with external fields\cite{klumpp_swimming_2019}, or with their environment \cite{bhattacharjee_bacterial_2019, bechinger_active_2016}. The latter is important as self-propelled particles often move in complex environments characterized by confining walls or interactions with obstacles. Many studies have investigated the microscopic origin of alignment with obstacles based on steric and/or hydrodynamic interactions\cite{wysocki_giant_2015, li_accumulation_2009, li_hydrodynamic_2014, Ostapenko18, hoeger_steric_2021}. Within a phenomenological description, such alignment can be implemented within ABP models as a 'wall torque', an empirical torque that provides the observed alignment with surfaces \cite{Ostapenko18, Codutti2022}. Experiments can inform some of the underlying parameters of such extended models, while others are strictly of an empirical nature, combining multiple physical phenomena.

Our goal in this study is two-fold. First, we demonstrate that the aforementioned empirical torque model is a quantitatively plausible description of an active rod's steric interaction with a surface and explore how its parametrization influences accuracy. For this, we utilize a classical least-squares approach to fitting the torque model to simulations of an active rod and evaluate the relevant residuals that describe the errors in the predicted dynamics. Second, we describe a procedure to learn the functional form of wall interactions, such as the wall torque from experimental data. This is done using a Bayesian inference approach, where the posterior model parameters are evaluated from trajectories of self-propelled particles. Notably, this approach does not require any information about the particle's orientation, which might be experimentally inaccessible.

\section{Model}

Our base model is an Active Brownian Particle (ABP)\cite{klumpp_swimming_2019}  in two dimensions with anisotropic diffusion, obeying the following set of Langevin equations:
\begin{align}
    \begin{split}
        \dot{\vect{r}}(t) &= v_0 \hat{\varphi} \vect{e}_{{x}} + \frac{1}{k_BT} \hat{\varphi}^{-1} \tens D_{\mathrm{T}} \hat{\varphi} \vect{F}_{\mathrm{wall}} + \sqrt{2} \hat{\varphi}^{-1} \sqrt{\tens D_{\mathrm{T}}}  \vect{\xi}(t), \\
        \dot{\varphi} (t) &= \frac{D_\mathrm{R}}{k_BT} \tau_{\mathrm{ABP}} + \sqrt{2D_{\mathrm{R}}} \chi(t). \\
    \end{split}
    \label{eq:ABP_EOM}
\end{align}
They describe the time evolution of the active particle's location $\mathbf{r}$ and its orientation $\mathbf{\varphi}$ using the active velocity $v_0$ and the rotational diffusion $D_R$. In contrast to the simplest ABP models, we allow for anisotropic translational diffusion, as we will use this model to describe elongated (rod-shaped) particles explicitly. The translational diffusion is thus described by a diagonal diffusion matrix $\tens D_{\mathrm{T}}=\mathrm{diag}(D_T^\parallel, D_T^\bot)$ in the particle's coordinate system. We use $\hat \varphi$ to denote the rotation matrix linking the lab's coordinate system to the particle's coordinate system. The diffusion is set to values representing a spherocylindrical rod, using the parametrization of 
Lüders et.~al.~\cite{luders_microscopic_2021}:
\begin{align}
    \begin{split}
        D_T^\parallel &= \frac{k_BT}{2 \pi \eta L}\left(\log (p) - 0.1404 + \frac{1.034}{p} - \frac{0.228}{p^2}\right), \\
        D_T^\bot &= \frac{k_BT}{4 \pi \eta L}\left(\log (p) +0.8369 + \frac{0.5551}{p} - \frac{0.06066}{p^2}\right), \\
        D_R &= \frac{3 k_BT}{\pi \eta L^3}\left(\log (p) - 0.3512 + \frac{0.7804}{p} - \frac{0.09801}{p^2}\right). \\
    \end{split}
    \label{diffterme_aniso}
\end{align}
$p$ is the aspect ratio of the spherocylinder, and $L$ is its length. This parametrization is valid for $p=1-30$ \cite{luders_microscopic_2021}.

The ABP may interact with walls, which exert force and torque on the particle.
The term $\vect{F}_{\mathrm{wall}}=-\nabla V_\mathrm{WCA}$ in eq.~\ref{eq:ABP_EOM} is the volume exclusion force exerted by walls. We use WCA repulsion \cite{WCA71}, a Lennard-Jones potential with a truncated attractive region:
\begin{equation}
    V_{\mathrm{WCA}}(x) =
    \begin{cases} 
    V_{\mathrm{LJ}}(x) - V_{\text{LJ}}(x_c) & \text{for } x < x_c =\frac{2^{1/6}}{2}\sigma, \\
    0 & \text{for } x \geq x_c,
    \end{cases}
\end{equation}
with
\begin{equation}
V_{\mathrm{LJ}}(x) = 4\epsilon \left[ \left( \frac{\sigma}{x} \right)^{12} - \left( \frac{\sigma}{x} \right)^{6} \right].
\end{equation}
Here, $\sigma$ is the length scale of the interaction. For our simulations, we set the hardness to $\epsilon=4k_BT$.

In addition, the ABP gets subjected to an empirical torque $\tau_{\mathrm{ABP}}$, factorized into a force term proportional to the instantaneous repulsion exerted by the wall (which depends on the distance from the wall) and an angle term dependent on the ABP's incidence angle with the wall:
\begin{equation}
    \tau_{\mathrm{ABP}} = \left|\vect{F}_{\mathrm{wall}}\right|\cdot f_\mathrm{ABP}\left(\theta_\mathrm{wall} - \varphi\right).
    \label{eq:full_torque_function}
\end{equation}
The angle term $f$ is an arbitrary function of the \emph{angle of incidence} between the swimming direction $\varphi$ and the wall's normal direction $\theta_\mathrm{wall}$. We aim to learn this function using information gathered from observations of an active rod's trajectory. For this, we parametrize the unknown function in terms of its spectral components
\begin{equation}
    f_\mathrm{ABP}(\varphi) = \sum_{\mathrm{n=1}}^{N_\mathrm{P}} \alpha_n
    \cdot \sin \left(n\left[ \theta_\mathrm{wall} - \varphi \right]\right).
    \label{eq:fitfunc_modes}
\end{equation}
up to order $N_\mathrm{P}$. All anti-symmetric cosine terms are suppressed by demanding symmetry with respect to the wall-normal, leaving only the sine terms. The factors $\alpha_i$ determine the modes' strengths. The first mode ($n=1$ with $\alpha_1<0$) always produces a torque away from the wall, irrespective of the angle of incidence. This creates an additional repulsion mechanism that leads to trajectories moving away from the wall in addition to the WCA repulsion. The second mode ($n=2$) creates a torque that aligns the ABP with the wall tangent. When the particle is heading toward the wall, the torque turns it away from the wall, whereas when the particle is headed away from the wall, the torque turns it back to the wall tangent. With additional higher frequencies, more complex torque functions can be described.

\subsection{Training Data generation}

\begin{figure*}
  \includegraphics[width=\textwidth]{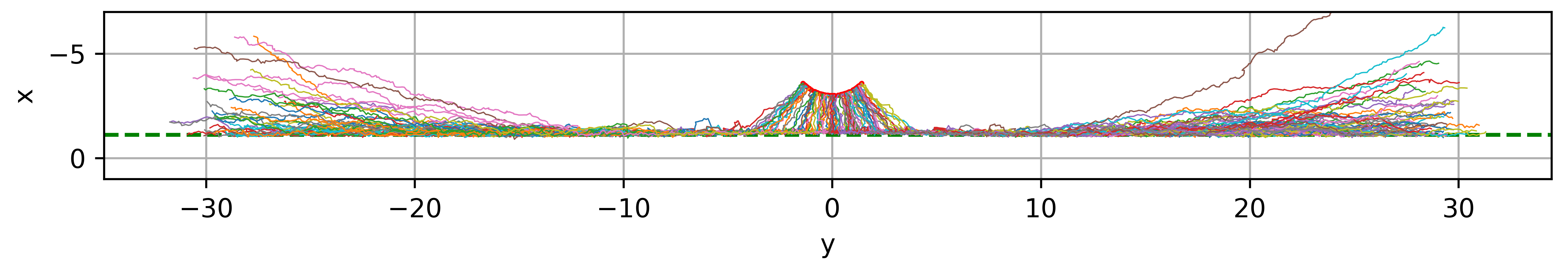}
  \caption{Synthetic observations generated for the analysis: Depicted are x- and y-coordinates of Type-R trajectories of single rods with $p=4$, colliding with a flat wall at $x=0\sigma$. The green dashed line denotes the cutoff distance of the WCA potential at $\left|x\right|=\frac{2^{1/6}}{2}\sigma$. All rods are initialized at (-5, 0). The traces are drawn using the rods' tips. \label{fig:single_collision}}
\end{figure*}

The Bayesian method outlined in this paper provides a framework for learning the wall torque function $f$ from experimental data on a swimmer's location over time. However, for the context of this paper, we only employ \emph{synthetic data}, as this allows for better diagnostics of the method. The data is generated using model variants described in the previous section. We generate two types of trajectories:
\begin{enumerate}
    \item \textbf{Type-A}: ABP trajectories with predefined wall torque functions $f$. These provide ground truth data to validate the Bayesian inference method.
    \item \textbf{Type-R}: Rod trajectories from a variant of our ABP model describing rod shape particles explicitly. In this case, the empirical wall torque is disabled, $f=0$. To capture steric interactions of a rod, the ABP gets equipped with $N_\mathrm{S}$ test sites equidistantly distributed at locations $\pm pi\sigma/2N_\mathrm{S}$ in the front and back of the ABP. The WCA repulsion forces $\vect F_\mathrm{\pm sat, i}$ are evaluated at the locations of the test sites and produce torques $pi\sigma/2N_\mathrm{S}\cdot\vect F_\mathrm{\pm sat, i}\times\hat\varphi\vect e_x$.
\end{enumerate}
Both types of simulations are set up like the example in fig.~\ref{fig:single_collision}. The ABP and the rods are initialized outside the WCA interaction region in front of the wall. The initial orientations were chosen on a grid from -60$^{\circ}$ to 60$^{\circ}$. Integration is performed using an Euler-Maruyama discretization and terminated when the particle is sufficiently separated from the wall. 

\section{Least-Squares Analysis}

To evaluate the performance of the empirical torque model in describing the steric interactions of rods, we've extracted the dynamic torque from type-R simulations.

\begin{figure}    \includegraphics[width=\columnwidth]{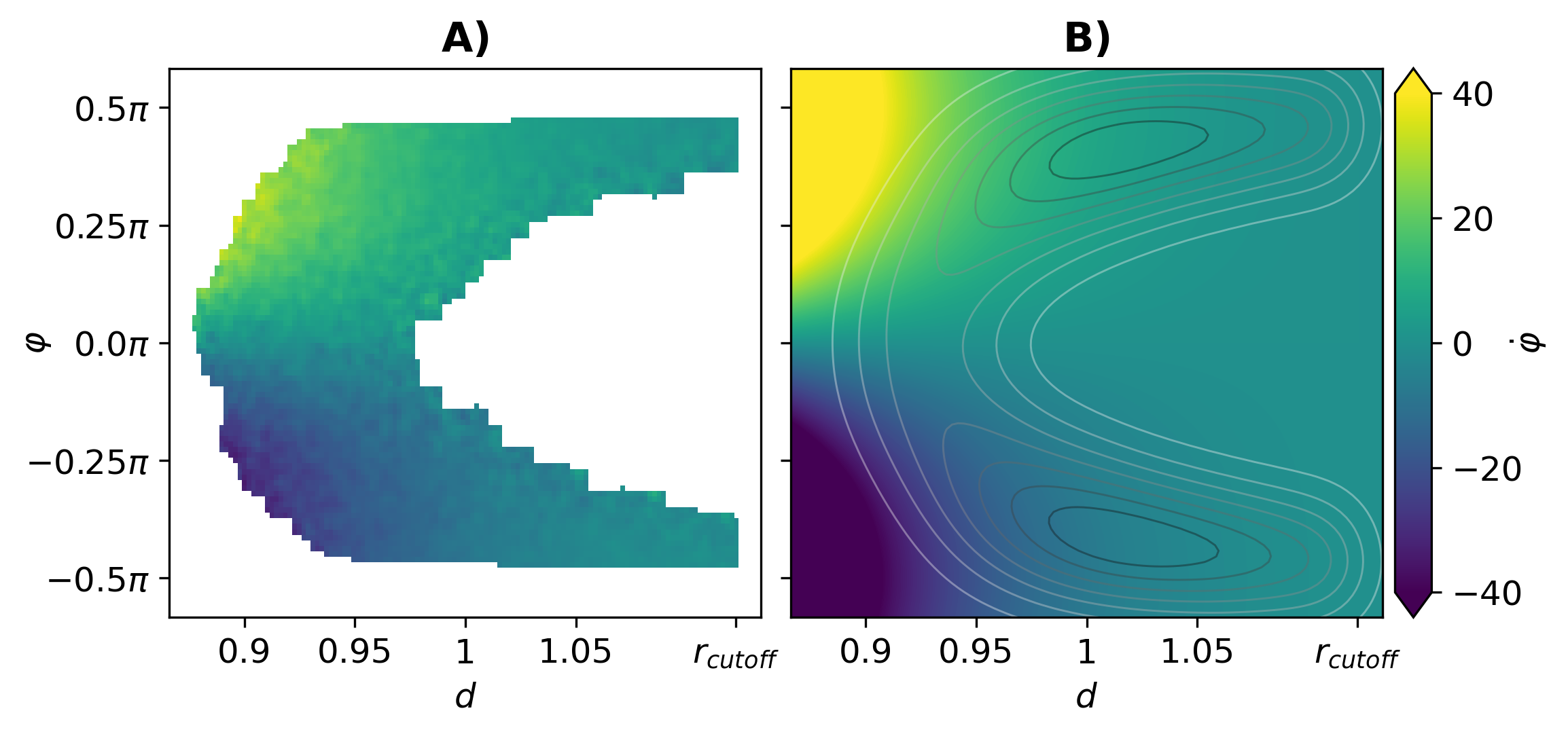}
    \caption{Scattering of a rod (Type-R) at a flat wall: Wall torque as a function of the distance $d$ from the wall (measured from the test site closest to the wall) and the orientation $\phi$ of the active rod. The aspect ratio was set to $p=5$. A: Conditional average of the torque as observed in Type-R simulations.  White regions were not visited by the scattering trajectories. B: Least-squares fit of a single-mode torque function $f$ with $n=1$ to the data from A. Grey lines indicate sample density collected from simulations.}
    \label{fig:example_fit}
\end{figure}

We generated $2500$ rod trajectories (100 per initial orientation), each for a range of parameter sets. The rate of change $\dot\varphi(t) \approx (\varphi(t+h) - \varphi(t))/h$ in the orientations, conditionally averaged for given distance and orientation, can then be used to calculate the effective torque on the rod:
\begin{equation}
    \tau(t) = \frac{k_BT}{D_{rot}}\, \langle\dot\varphi(t)\rangle.
\end{equation}
Fig.~\ref{fig:example_fit}a shows an example of the resulting torque data, parametrized by the distance to the wall and the rod orientation. The system's dynamics primarily occur in a crescent moon-shaped region, where the rod is inbound to the wall. The white regions are not sampled in the simulations, i.e., they are typically not visited during the scattering process. When the rod is pointed away from the wall, the torque quickly vanishes as the active motion and the repulsive interaction move it out of the interaction zone of the WCA potential.

We then fit eq.~\ref{eq:full_torque_function} to the data by minimizing the squared residuals:
\begin{equation}
    (\alpha_n)_{\mathrm{opt}} =\underset{(\alpha_n)}{\arg\min}\sum_{t,k}\left[ \tau^{(\alpha_n)}\left(\varphi^k(t), d^k(t)\right) - \tau^k(t)\right]^2,
\end{equation}
where the index $k$ runs over the simulated trajectories.
Assuming Gaussian errors in the calculated torques (which is expected given the diffusive structure of the equations of motion), this gives us a point estimate of the most likely torque parametrization. We calculate error bands using 68\% confidence intervals obtained from bootstrapping \cite{efron_bootstrap_79}. Fig.~\ref{fig:example_fit}B shows the best fit obtained for the leading mode ($n=1$). The corresponding amplitude is  $(\alpha_1)_{\mathrm{opt}}=2.03\sigma\pm0.08\sigma$.

We've tested the influence of several parameters on the model. Variations in the number of test sites for the repulsive interaction with the wall, the Péclet number of the active rod, and the wall hardness did not significantly influence the performance of the empirical model or the parametrization of $f$ (see Supplementary Material).

\begin{figure}
    \centering
        \includegraphics[width=1.0\linewidth]{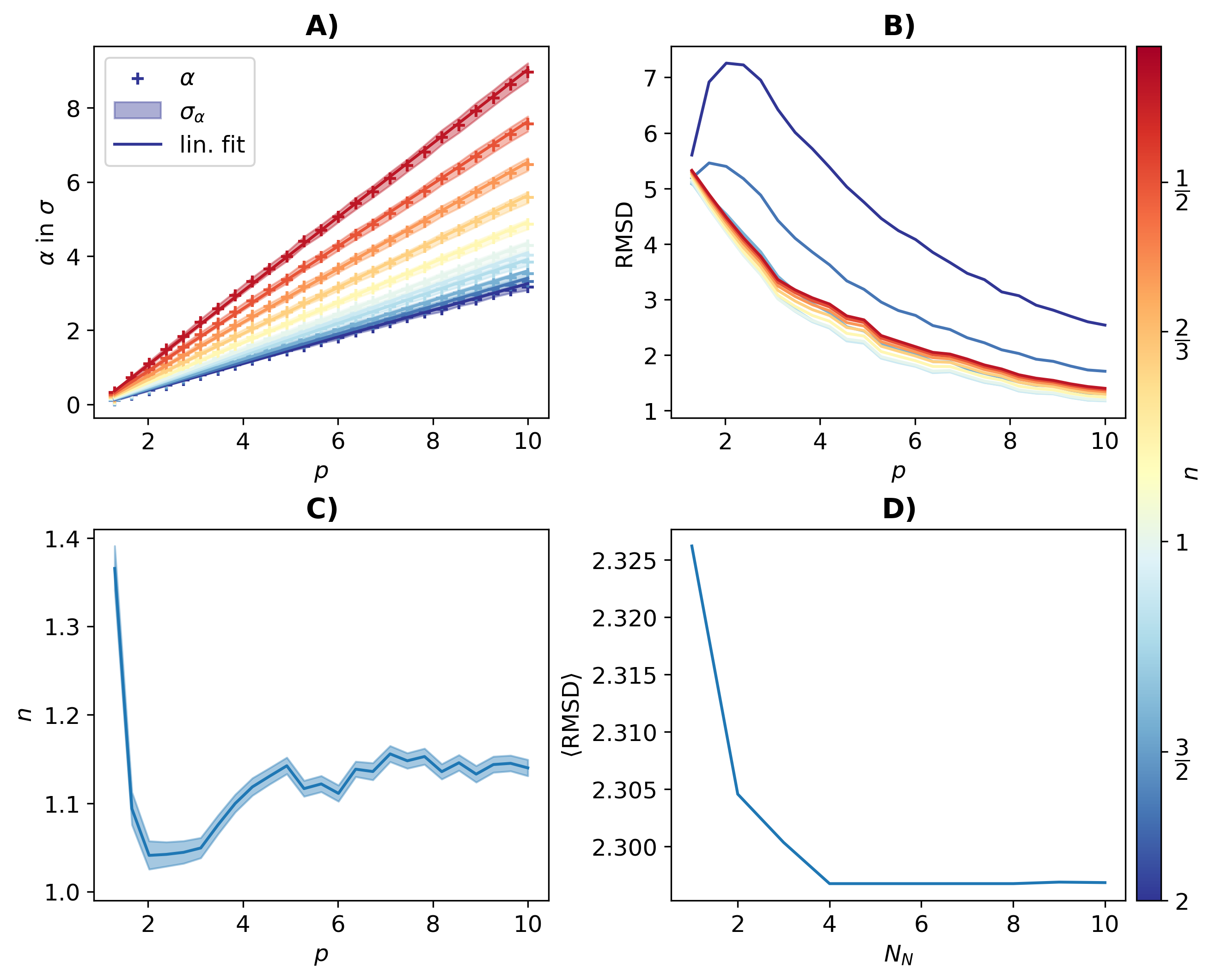}
    \caption{A) Optimal amplitude $(\alpha_n)_{\mathrm{opt}}$ and B) minimized residuals of the fit results as functions of the aspect ratio $p$. The color of the lines encodes the mode frequency $n$, with 68\% confidence bands. C) Best performing frequency $n$ as measured by RMSD. D) Model performance for an optimized model with $N_N$ modes, averaged over aspect ratios $p=1.5...10$.}
    \label{fig:aspect_ratio_data}
\end{figure}

In contrast, the rod aspect ratio $p$, the mode frequency $n$, and the mode count $N_\mathrm{P}$ affect the fit significantly. Fig.~\ref{fig:aspect_ratio_data}A and B show the fit results and performance for a single-mode description of $f$. Note that we've varied $n$ continuously instead of only considering integer values. We find a linear relationship between $1/n$ and its optimal amplitude $(\alpha_n)_{\mathrm{opt}}$. This suggests that the rod dynamics at small angles of incidence, where $f\propto\alpha_nn$, are of primary importance to the structure of $f$.
Furthermore, we see a linear relationship between the amplitude and the rod aspect ratio, corresponding to a modulation of the lever arm length. Surprisingly, the tangentially aligning mode ($n=2$) is not the optimal configuration. Instead, we found that a slightly bigger $n\simeq 1.1$ shows lower residual errors for all aspect ratios, as shown in fig.~\ref{fig:aspect_ratio_data}C.

When using multiple modes (discrete $n=1,2,...$), we find very similar results to the single-mode case in terms of performance and dependency on the aspect ratio. The ratios of optimal mode amplitudes show a non-linear dependency on $p$ (see Supplementary Material).  
Using up to four modes improves the model's performance, albeit by less than 1\% compared to the single-mode model. We did not find that using more than four modes leads to any further improvement of the model (see fig.~\ref{fig:aspect_ratio_data}D).

From these results, we conclude that using the empirical torque function $f$ successfully introduces a mechanism that captures the steric repulsion seen in rod simulations. However, different modes of the torque function result in approximately equal performance, indicating that the choice of torque function is not unique and that modes are somewhat interchangeable. Likewise, extending the representation to include higher-frequency components does not significantly improve the model quality after four modes. 

\section{Bayesian Analysis}

We now turn to a Bayesian approach \cite{BayesInPhysics, Murray2015} to learn the torque function $f$ from observations. This approach provides intuitive and easily interpretable \cite{Morey2015} results that quantify our knowledge about $f$ that is gained by observing the system dynamics. Specifically, we construct the posterior probability distribution of $f$. 

\subsection{Method Outline}

Bayesian inference targets the posterior distribution $p(\theta|Y)$, where $\theta$ denotes the set of parameters of interest, in our case, the parametrization $(\alpha_n)$ of the torque function $f$, and $Y$ denotes the data used for the inference, here consisting of sequences of $x$- and $y$-positions that form the trajectories of the rod (but not its orientation). We use the tip of the rod as the measured location, as it is the primary interaction point of the rod with the wall. The posterior distribution is calculated from Bayes' rule $p(\theta|Y) \propto p(Y|\theta)p(\theta)$, using the likelihood $p(Y|\theta)$ and the prior $p(\theta)$. We use an uninformative uniform prior for all components of $\theta$, only limiting the interval so that the ABP does not turn more than 90$^{\circ}$ in a single integration time step. Furthermore, we assume a flat prior for all unobserved orientational states.

The likelihood $p(Y|\theta)$ is computed for the (type-A) ABP model. Here, we exploit that our observation data originates from another simulation (type-A for test cases, then eventually type-R) with a finite integration time step. We adopt the same time steps for the ABP, thus avoiding introducing another level of discretization error. If the method is applied to real experimental data, one will introduce a discretization error corresponding to the data frequency (for example, the framerate of the experimental video). Using a single time step between observations conveniently decouples the rotational and translational diffusion of the ABP, making it possible to calculate the likelihood just from the particle's location while marginalizing the ABP's orientation with a filtering approach.

After discretization with an Euler-Maruyama scheme, equation \ref{eq:ABP_EOM} becomes
\begin{align}
    \begin{split}
    \vect{r}_{i+1} - \vect{r}_{i} - v_0 \hat{\tens\varphi}_i \vect{e}_{{x}}\Delta t - \frac{1}{k_BT} \hat{\varphi}^{-1}_i \hat{D}_{\mathrm{T}} \hat{\varphi}_i \vect{F}_{\mathrm{wall}}\Delta t\\
        =\sqrt{2} \hat{\varphi}_i \sqrt{\hat{D}_{\mathrm{T}}\Delta t}  \vect{w}_i(t) \equiv \vect{o}_i^{\varphi_i},
    \end{split}
    \label{eq:ABP_EOM_discrete}
\end{align}
yielding the displacement the particle experiences through translational diffusion as the offset $\vect{o}_i^{\varphi_i}$. The random numbers $\vect{w}_i$ are normal-distributed. Therefore, the offset is normal-distributed as well:
\begin{equation}
    \vect{o}_i^{\varphi_i} \sim \mathcal{N}(0,2\hat{\varphi}_i\hat{D}_{\mathrm{T}}\hat{\varphi}_i^{-1}\Delta t).
\end{equation}
Then, the joint likelihood of all the offsets, conditioned on the orientations $\varphi_i$ of the ABP at each time step, is
\begin{equation}
    p(\vect{o}_1^{\varphi_1}...\vect{o}_N^{\varphi_N}|\theta, \varphi_1...\varphi_N)=\prod_i^N\mathcal{N}(\vect{o}_i^{\varphi_i}|0,2\hat{\varphi}_i\hat{D}_{\mathrm{T}}\hat{\varphi}_i^{-1}\Delta t)
    \label{eq:offsets_likelihood_conditional}
\end{equation}

\subsection{State Marginalization}

This equation needs to be marginalized for all orientations $\varphi_i$, which is not analytically possible due to the trigonometric functions in $\hat\varphi_i$. As we're dealing with a state space model, an effective strategy is using a sequential filter, which calculates the posterior of $\varphi_i$ given all previous states $\varphi_1...\varphi_{i-1}$. The filter then steps through all time steps in sequence to construct the joint probability distribution $p(\varphi_1...\varphi_N|\theta)$ iteratively. We considered a range of commonly used filters and found that particle filters \cite{Andrieu2010}  work very well, unlike linear Kalman filters. We also decided against non-linear extensions of the Kalman filter as the particle filter produces an unbiased estimation of the posterior density, which allows us to sample from the true posterior via pseudo-Marginal Metropolis-Hastings (PMMH) methods \cite{pmmh_2009}.

The particle filter approximates the posterior distribution of the rotational state $\varphi_i$ as a sum over $k=1...N_p$ delta distributions at locations $\tilde\varphi_i^k$, the name-giving 'particles':
\begin{equation}
    p(\varphi_i | \theta, \varphi_1...\varphi_{i-1}) \approx \frac{1}{N_p}\sum_{k=1}^{N_p} \delta(\varphi_i - \tilde\varphi^k_{i-1})
\end{equation}
Filter particles, the discretization samples for the orientational states,  are characterized only by the unobserved orientational degree of freedom of ABPs, but, together with the positional data, can be used to sample full ABP trajectories that include the unobserved latent orientations. They follow the orientation dynamics of eq.~\ref{eq:ABP_EOM} with the torques that follow from the position data.
With this, we can marginalize equation \ref{eq:offsets_likelihood_conditional} as
\begin{align}
    p(Y|\theta) &\propto \int p(\vect{o}_1^{\varphi_1}...\vect{o}_N^{\varphi_N}|\theta, \varphi_1...\varphi_N)p(\varphi_1...\varphi_N)\upd\varphi_1...\upd\varphi_N \nonumber\\
    &\approx \prod_{i=0}^N\frac{1}{N_\mathrm{p}}\sum_{k=1}^{N_p} p(\vect{o}_i^{\tilde\varphi_i^k}|\theta, \varphi_1^k...\varphi_{i-1}^k)
    \label{eq:offsets_likelihood_marginalized}
\end{align}
Note that the offsets are the only stochastic terms in $Y$ when conditioned on the latent orientations, which is why, inside the integral, the offsets form the entirety of the observed data. The particles' initial states are sampled from the prior $\tilde\varphi^k_0 \sim p(\varphi_1)$. After that, the particles follow the orientational dynamics of equation \ref{eq:ABP_EOM} and are resampled with weights $w_i^k=p(\vect{o}_i^{\tilde\varphi_i^k}|\theta, \varphi_1^k...\varphi_{i-1}^k)p(\varphi_{i-1})$ after each time step. We do the resampling step before applying rotational diffusion to reduce particle degeneracy, where the orientation distribution is poorly represented as many identically resampled particles.

To summarize, we employ the following procedure to construct an estimation of the posterior density of the parameter set $\theta$:

\begin{enumerate}
    \item Initialize $N_\mathrm{p}$ filter particles (discrete orientation values) sampled from the prior $p(\varphi_0)$.
    \item Likelihood estimation: Calculate the offsets $\vect{o}_i^{\tilde\varphi_i^k}$ for all particles and update the partial product in equation \ref{eq:offsets_likelihood_marginalized}.
    \item Assimilation: Resample the particles based on the latest likelihood estimation and the priors on the orientations.
    \item Dynamics: Apply the equation of motion \ref{eq:ABP_EOM} to all filter particles, i.e. update the orientation by rotational diffusion and the torques.
    \item Repeat from 2 until all data has been used.
\end{enumerate}

\subsection{Posterior Density calculation}

The procedure described above does not normalize the posterior density, as the marginal probability is intractable to calculate. We address this with two different techniques. The first is to use a Markov-Chain Monte-Carlo (MCMC) method, the PMHH algorithm \cite{pmmh_2009},  to sample directly from the posterior. This method relies on the particle filter as an unbiased estimator of the posterior density. For visualizations of the posterior, we used the second approach, discretizing the parameter space $\theta$ and doing an exhaustive sweep. The discretization is then readily normalized.

\begin{figure*}
    \centering
        \includegraphics[width=1.0\linewidth]{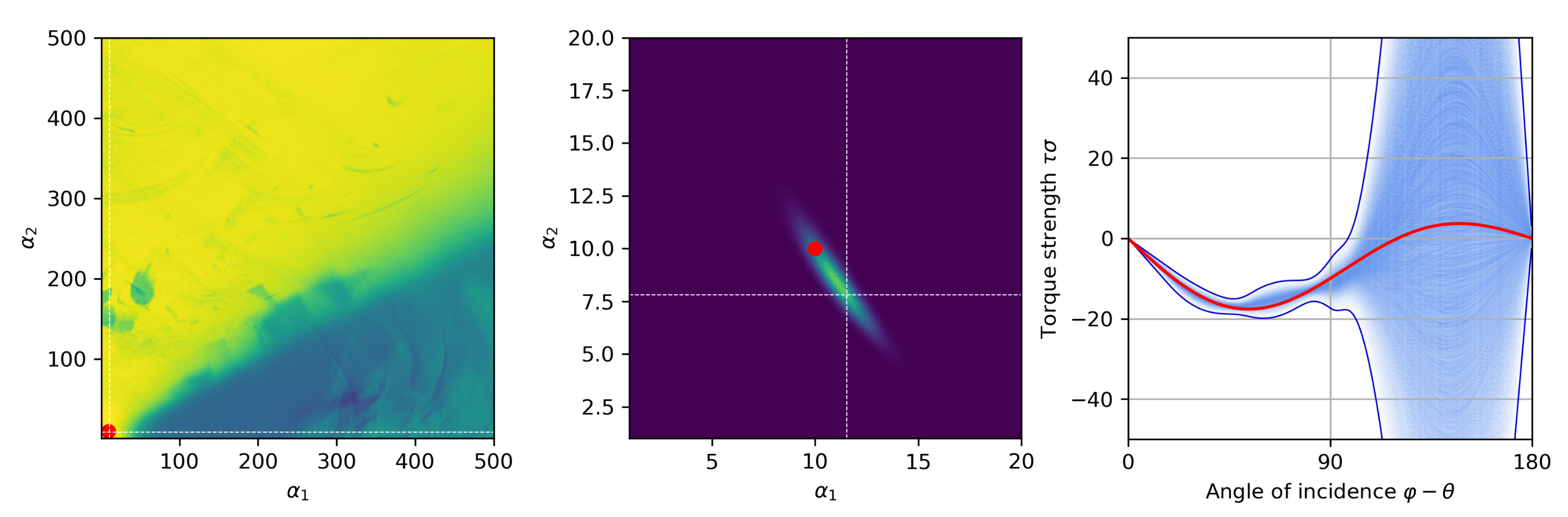}
    \caption{Test case for learning the posterior torque function $f$: Bayesian inference from 20 Type-A ABP trajectories generated with a known torque function $f$ (highlighted in red), parametrized with spectral parameters $\alpha_1 = \alpha_2 = 10$. A) The structure of the posterior space is drawn with logarithmic scaling of the density to make the complex structure visible. The data is obtained from a parameter sweep. Here, the parametrization is the same as the ground truth model. The dashed lines highlight the maximum of the distribution. B) Zoomed-in view of the bulk of the posterior mass, with linear mapping from density to color. C) Posterior of the torque function with the ground truth drawn in red. Here, the inference space is a 4D parametrization using B-splines, not a spectral representation like in A and B.
    }
    \label{fig:bayes_two_modes}
\end{figure*}

\subsection{Posterior Envelope}
To tune the inference performance in large parameter spaces, we employ a procedure to adapt a multivariate Gaussian envelope to the posterior. This can be done with relatively few filter evaluations, and the resulting distribution is then used directly as a proposal for the Markov chain.

Our fitting procedure produces a sequence of Gaussians $\mathcal{N}(\vect{\theta}_j, \tens{\Sigma}_j)$. Starting with an initial guess $\vect{\theta}_0$, $\tens{\Sigma}_0$, we draw $k=1...N_T$ parameter vectors from a test distribution $\vect{\theta}_j^k \sim \mathcal{N}(\vect{\theta}_j, z\tens{\Sigma}_j)$, with the windows size $z$ serving as a scaling of the search window. We then estimate the posterior densities $p(Y|\vect{\theta}_j^k)$ at the test points using the particle filter and calculate the next iteration as
\begin{align}
    \vect{\theta}_{j+1} &= \Sigma_{k=1}^{N_T}w_j^k\vect{\theta}_j^k\\
    \tens{\Sigma}_{j+1} &= \sum_{k=1}^{N_T} w_j^k \left(\vect{\theta}_j^k - \vect{\theta}_{j+1}\right)^T \left(\vect{\theta}_j^k - \vect{\theta}_{j+1}\right),
\end{align}
with weights
\begin{equation}
    \tilde{w}_j^k = \frac{p(Y|\vect\theta_j^k)}{\mathcal{N}(\vect\theta_j^k|\vect\theta_j, z\tens{\Sigma}_j)} \quad w_j^k = \frac{\tilde{w}_j^k}{\sum_{i=1}^{N_T}\tilde{w}_j^k}.
\end{equation}

This weighs test samples based on the targeted posterior density and compensates for the test distribution, which biases the sampling towards its mean $\vect\theta_j$. This procedure is iterated until convergence is observed in the covariance matrix. Convergence can be monitored from the entries of $\tens\Sigma_j$ (see Supplementary Material). We've found that $N_T=1024$ test samples, 1500 particles, and a search radius of $z=1.5$ work well in our scenarios, but we have not tuned the procedure in detail.

\subsection{B-Spline representation}

In our tests, we've found that using the spectral representation from eq.~\ref{eq:fitfunc_modes} is causing numerical issues when using more than 3-4 modes. In particular, the posterior mass collapses to a very thin distribution that is not aligned with any specific $\alpha_i$ dimension. The strong correlation makes it numerically difficult to find and represent the envelope of the posterior accurately, given the noisy density estimates from the filter. We attribute this to the fact that the Fourier modes are highly non-local and are affected by all observations.  By switching to a quadratic B-spline representation, we can alleviate the problem. The B-spline is of the form
\begin{align}
B^{(2)}(x) &= \sum_{i=0}^{N_B}\alpha_i\cdot\kappa\left((N-3)x + i - \frac{1}{2}\right)\\
\tx{ with }\kappa(u)&= \begin{cases}
\frac{1}{2}u^2                         & 0\le u \le 1,\\
\frac{1}{2}(-2u^2 + 6u - 3)& 1\le u \le 2,\\
\frac{1}{2}(3 - u)^2  & 2\le u \le 3, \\
0 & \tx{else}
\end{cases}
\end{align}
and represents functions in the base system $\kappa(.)$, stitched together from quadratic functions. The most important property of this spline for our use case is that $\kappa$ has compact support and thus is not affected by all observations, like the spectral representation.

\subsection{Validation on type-A data}
We tested the procedure on data generated from (Type-A) ABP simulations with known torque function $f$. An example is shown in fig.~\ref{fig:bayes_two_modes}A and B. We used a 2-mode torque (with ground truth $\alpha_1 = \alpha_2 = 10\sigma$) and generated 20 trajectories. We then inferred the amplitudes to check if the implementation correctly recovers the ground truth. The posterior density shows a highly complex structure that depends on the particular observations used for the inference. Most of the probability mass concentrates into a Gaussian near the ground truth. In all our tests, the 99\% highest density interval (HDI) reliably captures the ground truth. One can see a clear anticorrelation between the amplitudes of the two modes, indicating that the modes are, to some extent, interchangeable.

We also tested the inference of a B-spline-represented torque function on the same data (fig.~\ref{fig:bayes_two_modes}C). This representation also reliably captures the torque function, at least for inbound trajectories. Strikingly, the torque function is not well-learned in the outbound orientations when the particle moves away from the wall. This matches the observations from the least-squares approach, where dynamical data is collected in a crescent-shaped region of the $\varphi-d-$space. Information about the torque function for a specific orientation is primarily gained when the particle is inside the WCA interaction region and pointing in the respective direction. As the center of the crescent shape is never realized in the dynamics, no information about the torque can be gained.

\subsection{Inference on rod data}

\begin{figure*}
    \centering
    \includegraphics[width=1.0\linewidth]{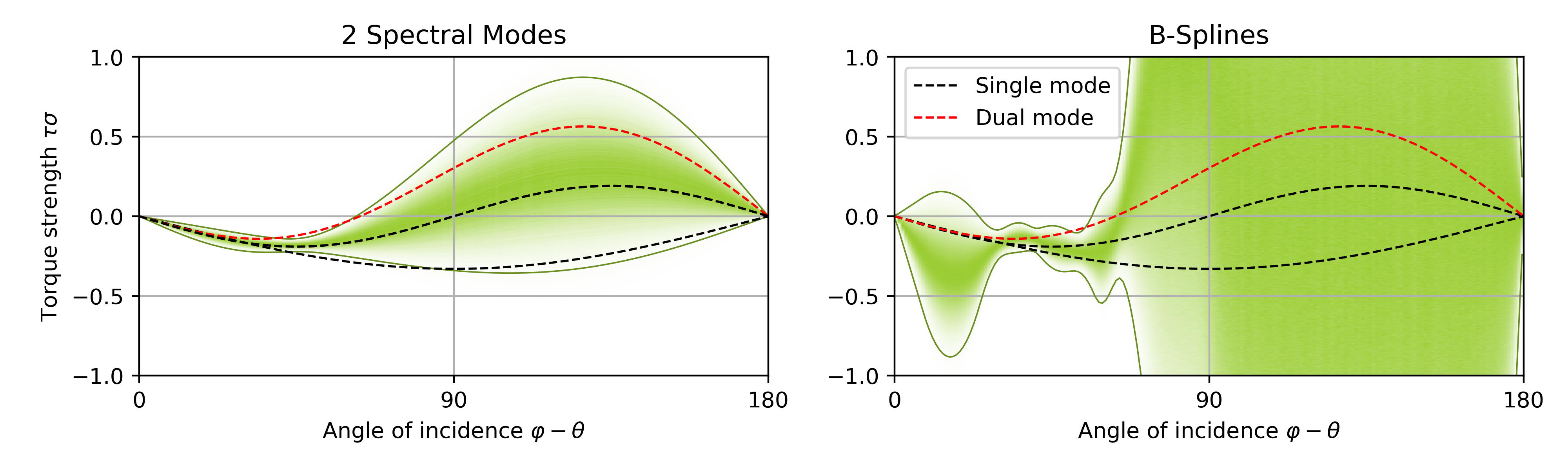}
    \caption{Posterior densities of function $f$ for an ABP tracking an active rods' tip. The posterior is parametrized using two-mode amplitudes on the left and 10 B-spline knots on the right. Inference is generated from 20 simulated trajectories of active rods (Type-R) that resemble data that can be obtained experimentally by tracking a swimmer. The green shaded area is the density of the posterior. It is enveloped by its 99\% HDI, indicated by green lines. The dashed lines are the result of the classical fitting approach using a single mode (black) and two modes (red).}
    \label{fig:rod_inference}
\end{figure*}

Finally, we learn the torque function $f$ from 20 Type-R simulations of active rods, which serve as a proxy for experimentally collected data. As these simulations now use true steric interactions with the wall instead of a predefined torque function, no ground truth for $f$ is known. Results for rods with aspect ratio $p=1.5$ are shown in fig.~\ref{fig:rod_inference}. Notably, the posterior density coincides with our fitting results for trajectories that are inbound to the wall, giving credence to the previous interpretation that this is the regime where the torque function is most important. The Bayesian inference furthermore reveals that most knowledge about $f$ is gathered in the inbound direction. The HDI remains broad for parallel and outbound trajectories.

Model selection imposes strong constraints on the torque function: Using two modes limits the bandwidth of the function, correlating the inbound and the outbound direction (fig.~\ref{fig:rod_inference}A). In this case, the HDI remains finite and relatively small (compared to the prior) for outbound movement, as knowledge about the regime is learned from information that is gathered from inbound movement.

Using a higher resolution representation of $f$, such as the B-splines representation shown in fig.~\ref{fig:rod_inference}B lifts these correlations. Then, the posterior $f$ densities for inbound and outbound movement are constructed from different observations. This reveals two important properties. The first is that $f$ is not strongly defined for small angles of incidence. While the inference strongly prefers negative torques (which corresponds to steric repulsion), there's no dominant magnitude that is needed to explain the trajectories. This may be explained by rotational diffusion being more important when a normally oriented rod experiences little torque, in combination with a lack of data on such rods, as they are in an unstable configuration. Secondly, the uncorrelated B-spline representation shows that the torque function does not impact the trajectory prediction in the outbound regime. The posterior coincides with the prior in this regime.

\section{Conclusions}

In this study, we have validated an empirical torque model to describe the steric interactions of active rods with walls. To this extent, we've simulated spherocylindrical active rods scattering with a wall and compared the resulting torques to an Active Brownian Particle model, which experiences a torque described by an orientation-dependent function $f$. By employing a least-squares approach, we demonstrated how the ABP model accurately fits the simulated rod trajectories. We've looked at the functional structure of $f$ and found that it is primarily influenced by the rod's aspect ratio. We've discovered no significant impact of the discretization of the rod or its activity on the model's accuracy.

We also introduced a Bayesian inference framework to learn the torque model from observational data. This method works robustly in our test cases, where we generated trajectories with predefined torque functions, which the inference could reasonably recover. The posterior density of the torque function directly highlights where information about the system is gained and where the model stays elusive. We've shown that care must be taken when selecting the representation of $f$, as insufficient complexity introduces unwarranted correlations that suppress details of the inference. We've applied the method to data obtained from rod simulations and found that the empirical torque model is effective for rods inbound to a wall, whereas the outbound direction is, in practice, not affected by the choice of torque model.

The Bayesian inference method presented here relies on the assumption that the system dynamics can be described by a single Euler-Maruyama step between observation times. While this is exact for our synthetic data, it will be an approximation when the method is applied to gather information from real experiments, and its accuracy remains to be explored. Finally, we want to emphasize that the Bayesian approach described here to infer wall torques is general and also applicable to inferring other properties of active particles. 

\acknowledgments
This work was supported by the Deutsche Forschungsgemeinschaft (DFG, German Research Foundation) – project ID 446142122. 
Simulations were run on the GoeGrid cluster at the University of G\"ottingen, which is supported by DFG (project IDs 436382789; 493420525) and MWK Niedersachsen (grant no.\ 45-10-19-F-02).

\bibliographystyle{eplbib}
\bibliography{bib}

\end{document}